\documentclass[12pt]{article} 
\RequirePackage[colorlinks,citecolor=blue,urlcolor=blue]{hyperref}
\usepackage[utf8]{inputenc} 

\usepackage{geometry} 
\usepackage{graphicx} 

\usepackage[dvipsnames]{xcolor}
\usepackage{booktabs} 
\usepackage{array} 
\usepackage{paralist} 
\usepackage{verbatim} 
\usepackage{subfig} 
\usepackage{amsbsy}
\usepackage{amsmath}
\usepackage{amssymb}
\usepackage{amsfonts}
\usepackage{amscd}
\usepackage{amsthm}
\usepackage{float}
\usepackage[toc,page]{appendix}
\usepackage{lineno}
\usepackage{mathrsfs}
\usepackage{todonotes}
\usepackage{setspace}
\usepackage{natbib}
\usepackage{multirow}
\usepackage{algorithm,algpseudocode}
\usepackage{varwidth}
\usepackage{ifthen}
\usepackage{pgfplots}
\usepackage{tikz}
\usetikzlibrary{decorations.pathreplacing,arrows,calc,positioning,automata,patterns,arrows.meta,math}
\usepackage{standalone}
\standaloneconfig{mode=buildnew} 
\usepackage{fancyhdr} 
\usepackage{rotating}
\usepackage[group-separator={,},group-minimum-digits=4]{siunitx}

\usepackage{sectsty}
\allsectionsfont{\sffamily\mdseries\upshape} 

\usepackage[nottoc,notlof,notlot]{tocbibind} 
\usepackage[titles,subfigure]{tocloft} 

\graphicspath{{img/}}

\newcommand\states{S}

\newcommand\numtips{N}

\newcommand{\substitutionModelParameters}{K}
\newcommand{\bigO}[1]{\mathcal{O}(#1)}

\begin{document}
\begin{flushright}
Article (Methods)\\
Version dated: \today\\
\end{flushright}

\bigskip
\medskip

\noindent{\Large \bf BEAGLE 4.1: A high-performance library for computation on phylogenetic trees across diverse parallel architectures}
\bigskip

\noindent{\normalsize \sc
	Karthik Gangavarapu$^{1}$,
	Xiang Ji$^{2}$,
	Yucai Shao$^{3}$,
	Philippe Lemey$^{4}$,
	Andrew Rambaut$^{5}$,
	Guy Baele$^{4}$,
	and Marc A.~Suchard$^{3,6,7,*}$ \vspace{1em} \\}
\noindent {\small
	\it $^1$Department of Translational Medicine, The Scripps Research Institute, La Jolla, California, United States \\
	\it $^2$Department of Statistics, College of Liberal Arts and Sciences, Iowa State University, Ames, IA, United States \\
	\it $^{3}$Department of Biostatistics, Jonathan and Karin Fielding School of Public Health, University of California, Los Angeles, United States \\
	\it $^4$Department of Microbiology, Immunology and Transplantation, Rega Institute, KU Leuven, Leuven, Belgium \\
	\it $^{5}$Institute of Ecology and Evolution, University of Edinburgh, Edinburgh, EH9 3FL, UK\\
	\it $^{6}$Department of Biomathematics, David Geffen School of Medicine at UCLA, University of 
	of California, Los Angeles, United States \\
	\it $^{7}$Department of Human Genetics, David Geffen School of Medicine at UCLA, University of California, Los Angeles, United States}
\\ \\
\noindent{\small
	\it * Corresponding author. E-mail: msuchard@ucla.edu}

\clearpage

\doublespacing
\paragraph{Abstract}

Efficient evaluation of sequence data likelihoods and their high-dimensional gradients on phylogenetic trees improves inference under both maximum-likelihood and Bayesian frameworks.
Here, we present BEAGLE 4.1, a high-performance library for statistical phylogenetics that incorporates new algorithms to evaluate these gradients on phylogenetic trees.
We also provide new hardware implementations for both likelihoods and gradients supporting ARM NEON intrinsics and optimized matrix multiplication units -- called tensor cores -- on NVIDIA graphics processing units (GPUs).
We benchmark the performance scaling of the library across a number of patterns and taxa on multi-core CPUs and GPUs, and  compare the speedup afforded by NVIDIA and AMD GPUs as well as performance scaling with an increasing number of GPUs.
We show that multi-core CPU implementations provide up to a fourfold speedup over single-threaded CPU implementations and up to an tenfold speedup for nucleotide and codon models, respectively, with performance generally improving as the number of taxa and site patterns increases.
GPUs outperform multi-threaded CPU implementations for a realistic number of patterns, even for nucleotide models with a small state-space size of 4, while for codon models they provide substantially higher performance gains even for a single pattern or four taxa.
Tensor cores on GPUs provide up to 2-fold speedup relative to standard CUDA cores for codon models.
Using NEON instructions on ARM CPUs affords up to a $\sim 1.3$-fold speedup over non-SIMD implementation with the speedup going down to 1.1-fold at 8 CPU threads.
Finally, we show that the CUDA implementation on NVIDIA devices is over 1.5-fold faster than the OpenCL implementation on a comparable AMD device and that using multiple GPUs can afford up to 4-fold higher performance over using a single GPU.
We provide these new algorithms to evaluate the gradient and efficient hardware implementations for both likelihood and gradient calculations through BEAGLE 4.1, such that they can be readily integrated into phylogenetic software packages.

\section{Introduction}

BEAGLE~\citep{ayres_beagle_2012,ayres_beagle_2019} is a high-performance computing library for statistical phylogenetics that supports parallelization on multi-core CPUs through multi-threading and fine-grained vectorization using Streaming SIMD Extensions (SSE) instructions, and on graphics processing units (GPUs) through CUDA~\citep{cuda2012} and OpenCL~\citep{stoneOpenCLParallelProgramming2010}.
BEAGLE is used by popular phylogenetic software such as BEAST X \citep{baele_beast_2025}, BEAST 2.5 \citep{beast2}, MrBayes \citep{Ronquist2012}, RevBayes \citep{smith2024}, and PhyML \citep{Guindon2010}.
Since the last version of BEAGLE~\citep{ayres_beagle_2019}, advances in algorithms and computing hardware offer new opportunities to accelerate phylogenetic inference.

Efforts to parallelize numerical calculations for statistical phylogenetics have largely focused on parallelizing Felsenstein's pruning algorithm \citep{felsenstein_evolutionary_1981} to calculate the likelihood on a phylogenetic tree of observed sequence data.
Phylogenetic inference is predominantly performed within maximum-likelihood and Bayesian frameworks, both of which benefit from efficient evaluation of the gradient of the log-likelihood with respect to (w.r.t.) model parameters.
Maximum-likelihood estimation relies on gradient-based optimization methods -- such as Broyden--Fletcher--Goldfarb--Shanno (BFGS) \citep{dennis1996numerical} -- to find parameter values that maximize the likelihood, while Bayesian inference benefits from efficient gradient-based Markov Chain Monte Carlo (MCMC) algorithms such as Hamiltonian Monte Carlo (HMC) \citep{neal2011mcmc} to sample from posterior distributions.

\begin{figure}[H]
	\centering
	\includegraphics[width=\textwidth]{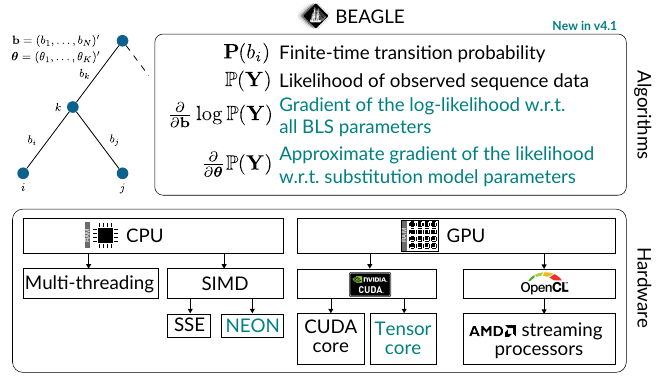}
	\caption{BEAGLE 4.1 is a high-performance library that accelerates numerical calculations for statistical phylogenetics with support for parallelization across CPUs and GPUs. To enable fine-scale parallelization on CPUs, BEAGLE supports SIMD instructions through SSE and NEON instructions. BEAGLE supports GPUs through the CUDA API for NVIDIA devices and OpenCL API for AMD devices.}
	\label{fig:beagleOverview}
\end{figure}

In this paper, we focus on algorithmic advances for efficiently evaluating the gradient of the log-likelihood w.r.t.\ branch-length-specific (BLS) parameters, such as branch-specific evolutionary rates \citep{ji_gradients_2020}, and substitution model parameters through an approximation of the gradient \citep{magee_random_effects_2024}.
In regard to hardware implementations, we target NEON SIMD instructions on ARM CPUs, whose adoption has grown steadily since the release of Apple Inc.\ M-series chips, and specialized matrix-multiplication cores available on modern NVIDIA GPUs for calculating the likelihood of observed sequence data on a phylogenetic tree and evaluate its gradient w.r.t.\ phylogenetic parameters.
We describe these advances in BEAGLE 4.1 (Figure~\ref{fig:beagleOverview}) which provides efficient hardware implementations for both calculating the likelihood and evaluating its gradient across heterogeneous computing hardware including multi-core CPUs and GPUs.

\section{Key Improvements}

In this section, we describe new algorithms and hardware implementations supported by the BEAGLE 4.1 library.

\subsection{Algorithmic advances}

\citet{ji_gradients_2020} introduce an algorithm to calculate the gradient of the log-likelihood w.r.t.\ all BLS parameters in linear time w.r.t.\ $\numtips$, the number of tips in a phylogenetic tree.
In a Bayesian framework, this algorithm enables inference using gradient-based HMC achieving 7- to 23-fold speedup in terms of median effective sample size (ESS) per second relative to random-walk proposals.
This algorithm also accelerates inference under a maximum-likelihood framework by 210- to 253-fold compared to evaluating the gradient numerically, which has a time complexity of $\bigO{\numtips^2}$.
The algorithm complements the post-order traversal of the tree with a pre-order traversal to calculate post- and pre-order partial likelihoods.
The likelihood of the observed sequence data is then calculated as the inner product of the post- and pre-order partial likelihoods, enabling evaluation of the gradient of the log-likelihood w.r.t.\ all BLS parameters in a single pre-order traversal using $\bigO{\numtips}$ operations.
This linear-time algorithm can also be used to simultaneously infer both the node heights and branch-specific evolutionary rates using HMC by applying a ratio transform on the node heights that inherently preserves topological constraints \citep{ji_scalable_2023}.

Random-effects substitution models extend standard continuous time Markov chain (CTMC) substitution models by introducing multiplicative random effect parameters in every non-diagonal element of the instantaneous rate matrix, enabling the model to capture deviations from standard substitution processes.
This substantially increases the number of parameters, making inference under these models computationally challenging.
\citet{magee_random_effects_2024} solve this problem by reformulating the gradient of the sequence likelihood w.r.t.\ all substitution model parameters as the product of the gradient w.r.t.\ elements of the rate matrix and the gradient of the matrix elements w.r.t.\ substitution model parameters, and exploiting a first-order approximation of the derivative of the matrix exponential to evaluate the former.
Approximating the gradient of sequence likelihood w.r.t.\ all substitution model parameters in this manner reduces the complexity from $\bigO{\substitutionModelParameters \numtips \states ^5}$ for a naive derivative to $\bigO{\substitutionModelParameters \states^2 + \numtips \states^3}$, where $\states$ is the state-space size and $\substitutionModelParameters$ is the number of substitution model parameters which can range from $1$ to $\states (\states - 1)$.
This allows for efficient inference under these models using HMC, yielding an average speedup of 6.6-fold for nucleotide models and 20.2-fold, in terms of ESS per second, for phylogeographic models with larger state-space sizes over standard random-walk proposals.

\subsection{Hardware advances}

\subsubsection{ARM NEON}

Starting with their M-series chips, Apple Inc.\ has transitioned to using ARM-based processors making them widely available in consumer computing and research environments.
ARM processors implement the NEON instruction set for SIMD operations while x86 processors use SSE.
Support for SSE was introduced in BEAGLE 3~\citep{ayres_beagle_2019}.
To enable vectorized operations on ARM processors, we used sse2neon.h \citep{sse2neon}, a header-only library, to translate the existing SSE instructions into their NEON equivalents.
By using this translation layer rather than maintaining two separate SSE and NEON implementations, we avoid duplicating our code while realizing the benefits of using SIMD operations on ARM processors.
Early versions of sse2neon.h lacked NEON mappings for a subset of double-precision SSE intrinsics such as \_mm\_div\_pd, \_mm\_store\_sd, and \_mm\_dp\_pd.
We initially provided manual implementations of these functions using native ARM NEON intrinsics and existing sse2neon.h utilities.
We retired these custom implementations as subsequent releases of sse2neon.h incorporated support for these instructions, allowing us to rely entirely on the library for SSE-to-NEON translation.

\subsubsection{Tensor cores}

Modern GPUs are equipped with specialized hardware units optimized for general matrix multiplication (GEMM) called tensor cores on NVIDIA devices.
Starting with the Ampere architecture~\citep{NVIDIAAmpere}, tensor cores have supported GEMM at double-precision floating-point format (FP64).
Tensor cores can be utilized for matrix-multiply-and-accumulate (MMA) operations using warp-level instructions provided in the CUDA API where a warp represents a group of 32 threads, the smallest unit of parallel execution on a GPU.
In recent work, we have shown that massively parallel, tensor core-based algorithms deliver 2 to 3-fold gains in performance for amino acid and codon models, accompanied by a $\sim 2$-fold reduction in energy usage for codon models, relative to using standard GPU-based algorithms that rely on CUDA cores \citep{gangavarapu2026}.
BEAGLE 4.1 supports tensor cores on NVIDIA devices to calculate the likelihood of observed sequence data and evaluate the gradient of the log-likelihood w.r.t.\ all BLS parameters.

\section{Performance evaluation}

\subsection{Scaling by number of taxa and site patterns}

\begin{figure}[H]
	\centering
	\includegraphics[height=0.8\textheight]{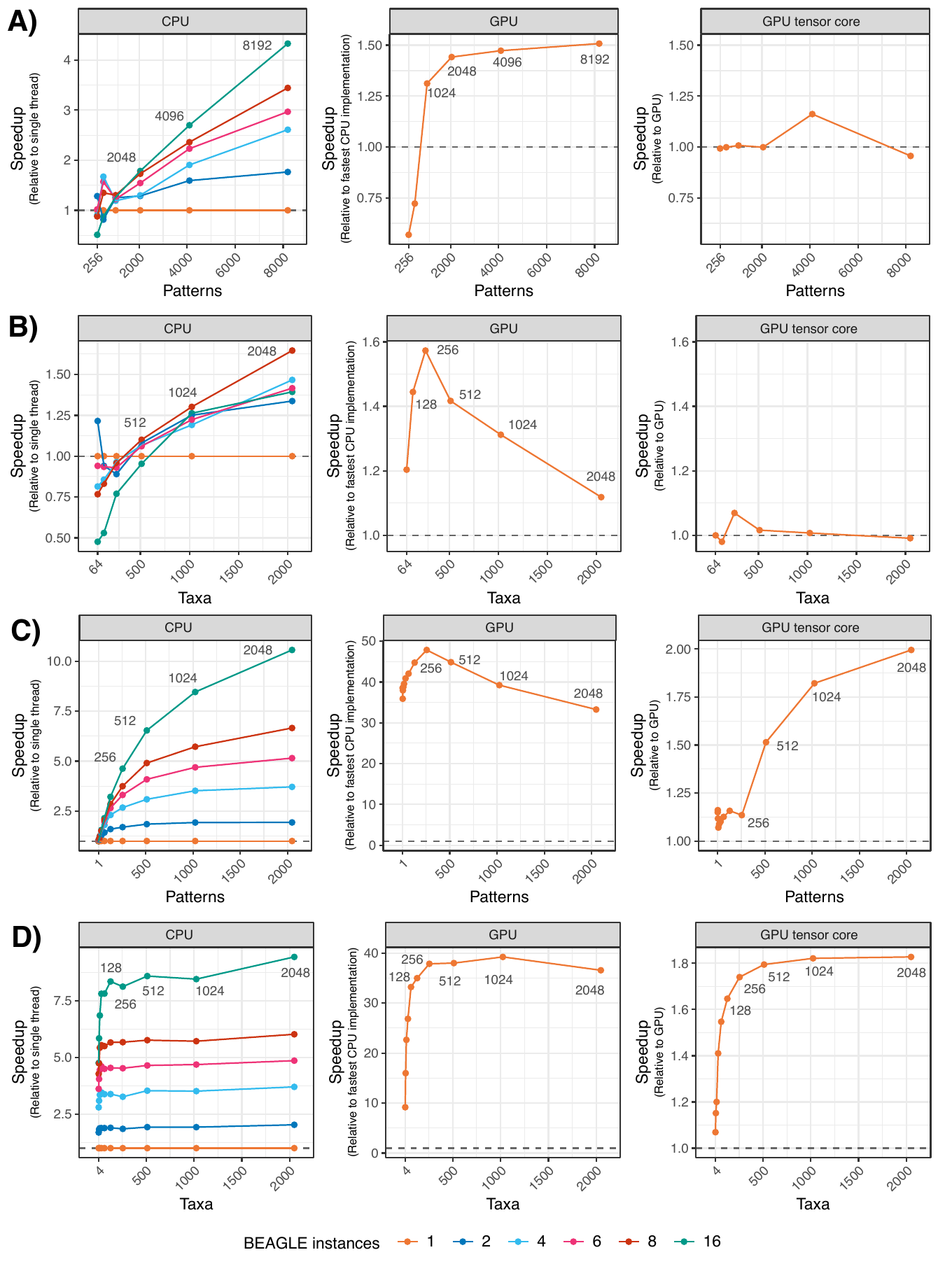}
	\caption{Multi-threaded CPU implementations outperform single-threaded CPU, GPUs deliver additional speedup over the fastest CPU implementation, and for codon substitution models, GPU tensor core implementation provides further gains over the standard GPU implementation. Speedups are reported as follows: multi-threaded CPU relative to single-threaded CPU, GPU relative to the fastest CPU implementation, and GPU tensor core relative to the standard GPU. Results are shown for a nucleotide substitution model as a function of (A) the number of patterns with fixed 1\,024 taxa and (B) the number of taxa with fixed 1\,024 patterns, and for a codon substitution model as a function of (C) the number of patterns with fixed 1\,024 taxa and (D) the number of taxa with fixed 1\,024 patterns.}
	\label{fig:speedUpBenchmarks}
\end{figure}

The number of unique site patterns and the number of taxa are the most important factors that affect the performance gains afforded by parallelized and massively-parallelized implementations in statistical phylogenetics.
We benchmarked the wall time of hardware implementations in BEAGLE 4.1 across these two factors by inferring branch-specific evolutionary rates given a fixed tree for two substitution models, a HKY nucleotide substitution model \citep{hasegawa_dating_1985} and a Goldman-Yang codon (GY94) model \citep{goldmanAndYang94}, on a system equipped with an AMD EPYC 9334 32-core processor and an NVIDIA H100 NVL GPU.
We measured speedup afforded by multi-core CPU and GPU implementations relative to a single-threaded CPU instance for 10 iterations of the MCMC using HMC over the evolutionary rates as implemented in BEAST X \citep{baele_beast_2025}.
We measured performance scaling by increasing the number of site patterns while keeping the number of taxa fixed at 1\,024, and by increasing the number of taxa while keeping the number of site patterns fixed at 1\,024. (Figure~\ref{fig:speedUpBenchmarks}).

For nucleotide models with a state-space size of 4, the speedup from multi-threaded CPU implementations increased with the number of unique site patterns with 16 threads, starting at 2\,048 patterns, achieving a maximum of 2- to 4-fold speedup over a single CPU thread (Figure~\ref{fig:speedUpBenchmarks}A).
Starting from 1\,024 patterns, the GPU implementation outperformed the fastest CPU implementation delivering a further 1.25- to 1.5-fold speedup.
The speedup from multi-threaded CPU implementations increased with the number of taxa, with 8 threads delivering speedups over 1.25-fold starting at 1\,024 patterns (Figure~\ref{fig:speedUpBenchmarks}B).
The CPU implementation using 8 threads outperformed 16 threads starting at 1\,024 patterns, indicating that the workload becomes memory-bound rather than compute-bound at this point.
The GPU implementation affords over 1.1-fold speedup over the fastest CPU implementation starting at 64 taxa.
We note that the performance of the GPU implementation saturates at 256 taxa and the diminishing speedup relative to the fastest CPU implementation reflects the growing speedup achieved by the 8-thread CPU implementation rather than a decline in GPU performance.
Using tensor cores on the GPU provided no additional performance gain over standard CUDA cores for nucleotide models which is expected given the small state-space size of 4.

For a codon model, adding more CPU threads consistently improved performance, with the 16-threaded implementation reaching over 10-fold speedup at 2\,048 patterns (Figure~\ref{fig:speedUpBenchmarks}C).
The GPU implementation afforded over 30-fold speedup relative to the fastest CPU implementation with 16 threads starting at just a single pattern.
We note that the performance of the GPU implementation saturates at 256 patterns and the diminishing speedup relative to the fastest CPU implementation reflects the growing speedup achieved by the 16-threaded CPU implementation rather than a decline in GPU performance.
Using tensor cores on GPUs for a codon model, with a larger state-space size of 61, delivered a further $\sim 1.25$-fold speedup over standard GPU implementation at 256 patterns reaching a maximum of 2-fold at 2\,048 patterns.
The CPU implementation using 16 threads consistently offered over 4-fold speedup starting at 4 taxa with performance saturating at $\sim 8$-fold speedup for 128 taxa (Figure~\ref{fig:speedUpBenchmarks}D).
The GPU implementation afforded $\sim 10$-fold speedup over the fastest 16-thread CPU implementation at 4 taxa saturating at $\sim 40$-fold speedup at 256 taxa.
GPU tensor cores provided additional speedup over the GPU implementation starting at $\sim 1.1$-fold speedup at 4 taxa, saturating at $\sim 1.8$-fold speedup starting with 512 taxa.
To demonstrate the speedup afforded by the GPU over CPU to infer parameters under a GY94 codon model with random effects at every non-diagonal element, we measured the total time spent in BEAGLE for 10 iterations of the MCMC using HMC over substitution model parameters.
We observed a 94-fold speedup when using the GPU relative to a single-threaded CPU implementation.

\subsection{ARM NEON}

We measured the speedup afforded by NEON SIMD instructions relative to a non-SIMD implementation on an M3 Pro ARM processor, using a dataset of 352 serotype 3 dengue viral genomes with 2\,294 unique site patterns.
We set up a phylogenetic model to infer branch-specific evolutionary rates with a fixed tree under a GTR nucleotide substitution model \citep{tavare1986lectures}.
We measured the wall-time for 10\,000 iterations of the MCMC with HMC over the branch-specific evolutionary rates as implemented in BEAST X across 10 replicates and an increasing number of threads.
NEON instructions afforded over 1.3-fold speedup on a single thread, and close to 1.1-fold speedup with 8 threads, compared to non-SIMD instructions for nucleotide substitution models with a state-space size of 4. (Figure~\ref{fig:speedup}).

\subsection{Benchmarks on NVIDIA and AMD GPUs}

\begin{figure}[H]
	\centering
	\includegraphics[width=\textwidth]{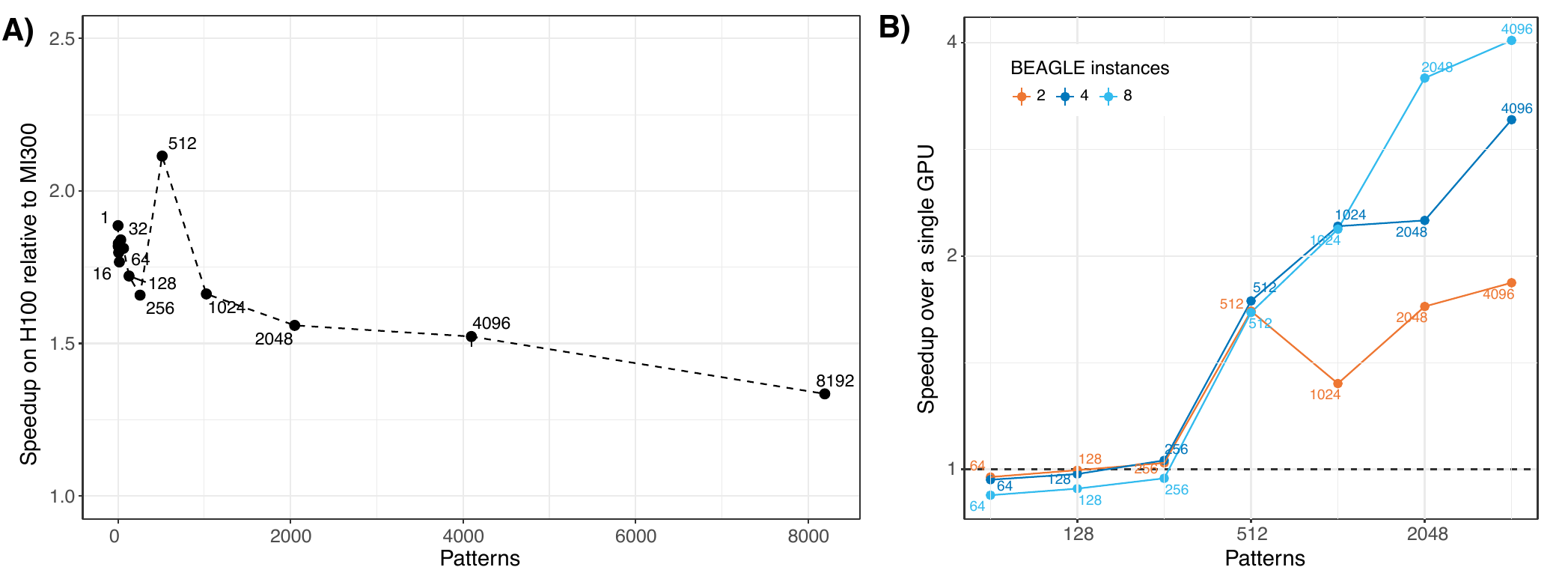}
	\caption{CUDA implementation on NVIDIA H100 delivers approximately 1.5-fold higher speedup than the one on AMD MI300X for codon models, and when using multiple GPUs, doubling the number of GPUs beyond 512 site patterns yields 1.1- to 1.7-fold incremental gain in speedup.
	(A) Speedup on an NVIDIA H100 relative to AMD MI300X for 10 iterations of MCMC for a codon substitution model across 10 replicates. (B) Speedup of using multiple GPUs with one BEAGLE instance per GPU over using a single GPU, measured on a system equipped with 8 AMD MI300X GPUs across 10 replicates.}
	\label{fig:gpuspeedup}
\end{figure}

BEAGLE 4.1 supports the two GPU providers NVIDIA and AMD through the CUDA and OpenCL APIs, respectively.
To compare the hardware-specific implementations, we set up a Bayesian phylogenetic analysis to infer branch-specific evolutionary rates for a dataset of 1\,024 taxa with an increasing number of patterns under a GY94 codon model with a fixed tree.
We measured the wall-time for 10 iterations of the MCMC with HMC over the branch-specific evolutionary rates as implemented in BEAST X on two comparable GPUs: AMD MI300X and NVIDIA H100 PCIe on cloud instances also equipped with comparable AMD EPYC 9474 and AMD EPYC 9454 CPUs, respectively.
In Figure~\ref{fig:gpuspeedup}A, we report the speedup afforded by CUDA implementation on the H100 over OpenCL implementation on the MI300X across ten replicates.
Across all pattern counts, the CUDA implementation delivered over 1.5-fold higher performance relative to the OpenCL implementation.

Systems equipped with multiple GPUs, typically up to 8, are becoming increasingly common in research environments.
BEAGLE supports multiple GPUs by partitioning the sequence alignment into conditionally independent blocks running on separate GPUs.
To measure the speedup afforded by utilizing multiple GPUs, we measured the wall time for 10 iterations of MCMC with HMC over branch-specific evolutionary rates using the same dataset with 1\,024 taxa under a GY94 codon model with a fixed tree.
We observe that using multiple GPUs results in performance gain starting with 512 patterns relative to using a single GPU.
Over 512 patterns, doubling the number of GPUs resulted in a 1.1- to 1.7-fold incremental gain in speedup (Figure~\ref{fig:gpuspeedup}B).

\section{Discussion}

BEAGLE 4.1 advances computational infrastructure for statistical phylogenetics by providing efficient hardware implementations to calculate the likelihood of observed sequence data and new algorithms to evaluate the gradient of the log-likelihood w.r.t.\ BLS and substitution model parameters.
While multi-threaded CPU implementations afford performance gains over a single thread, GPUs consistently outperformed CPU implementations for a realistic number of site patterns, even for nucleotide models with small state-space sizes.
Tensor cores on NVIDIA GPUs afford up to 2-fold speedup relative to standard CUDA cores for codon models with larger state-space size.

In future work, we plan to improve the caching strategies during post- and pre-order traversals of the tree.
For substitution models with several site patterns, transition probability matrices are calculated through numerical eigendecomposition of the infinitesimal rate matrix and cached in memory for reuse across patterns.
However, for phylogeographic problems that typically have a large state-space size but only a single site pattern, the transition probability matrices are not reused, and it is more efficient to form the transition probability matrix on-the-fly while calculating the likelihood of observed sequence data in this case through Felsenstein's pruning algorithm.
The linear-time gradient w.r.t.\ all BLS parameters is evaluated in a single pre-order traversal, through a sandwich operation of a pre-order partial likelihood vector, the infinitesimal rate matrix, and a post-order partial likelihood vector.
Currently, the post-order partial likelihood vectors and infinitesimal rate matrices are cached separately.
Since the infinitesimal rate matrix is already available during the post-order traversal, the product of the infinitesimal rate matrix and the post-order partial likelihood vector at each node could be cached directly thus reducing memory burden from $\states^2$ elements of the infinitesimal rate matrix to $\states$ partial likelihoods.
Finally, reverse-mode gradients involving the adjoint of the matrix exponential \citep{monti2025nonparametric} suggest exact substitution parameter gradients on trees in only $\bigO{\states^3 + \numtips \states^2}$ that we plan to introduce into BEAGLE.

BEAGLE 4.1 currently supports AMD GPUs through OpenCL, which does not provide access to AMD matrix cores which are optimized matrix multiplication cores analogous to tensor cores on NVIDIA devices.
In future work, we plan to add support for AMD devices through a Radeon Open Compute (ROCm) implementation in BEAGLE, which would enable use of AMD matrix cores for accelerating GEMM operations.
A second direction for future work is supporting inference at single precision, as commercial-grade GPUs, unlike data center GPUs, provide substantially higher performance at single-precision compared to double-precision floating point format.
BEAGLE is developed independent of tree-inference packages and hence, existing phylogenetic models can immediately take advantage of the advances described in this paper without requiring any explicit modifications to the models themselves.

\section{Data Availability}

Log files, XML files for BEAST X, and associated scripts to reproduce the results and figures have been deposited at \url{https://doi.org/10.5281/zenodo.20404492}.
The BEAST XMLs and instructions to replicate the benchmarks are available at \url{https://github.com/suchard-group/beagle4_supplement}.

\section{Software Availability}

BEAGLE 4.1 is available at \url{https://github.com/beagle-dev/beagle-lib/tree/master}.
BEAST X is available at \url{https://github.com/beast-dev/beast-mcmc/tree/main}.
We used compile-time constants to measure the time spent in BEAGLE and we have made this BEAST X JAR file available at \url{https://github.com/suchard-group/beagle4_supplement}.

\section{Acknowledgments}

KG is partially supported by US National Institutes of Health (NIH) grant U19 AI135995. 
PL acknowledges support by the Research Foundation - Flanders (‘Fonds voor Wetenschappelijk Onderzoek - Vlaanderen’, G010326N, G005323N and G051322N).
GB acknowledges support from the Research Foundation - Flanders (``Fonds voor Wetenschappelijk Onderzoek - Vlaanderen,'' G098321N), from the European Union Horizon 2023 RIA project LEAPS (grant agreement no. 101094685), and from the DURABLE EU4Health project 02/2023-01/2027, which is co-funded by the European Union (call EU4H-2021-PJ4) under Grant Agreement No. 101102733.
XJ, YS and MAS are supported by NIH grants R01 AI153044 and R01 AI162611.
This work was supported in part by Advanced Micro Devices, Inc. under the AMD University Program's AI \& HPC Cluster.

\bibliographystyle{biometrika}
\bibliography{beagle4_manuscript}

\newcommand{\beginsupplement}{%
		\setcounter{table}{0}
		\renewcommand{\thetable}{S\arabic{table}}%
		\setcounter{figure}{0}
		\renewcommand{\thefigure}{S\arabic{figure}}%
		\setcounter{section}{0}
		\renewcommand \thesection{S\arabic{section}}
		\setcounter{algorithm}{0}
		\renewcommand{\thealgorithm}{S\arabic{algorithm}}
	}

\beginsupplement
\section{Supplementary Material}

\begin{figure}[H]
	\centering
	\includegraphics[width=0.75\textwidth]{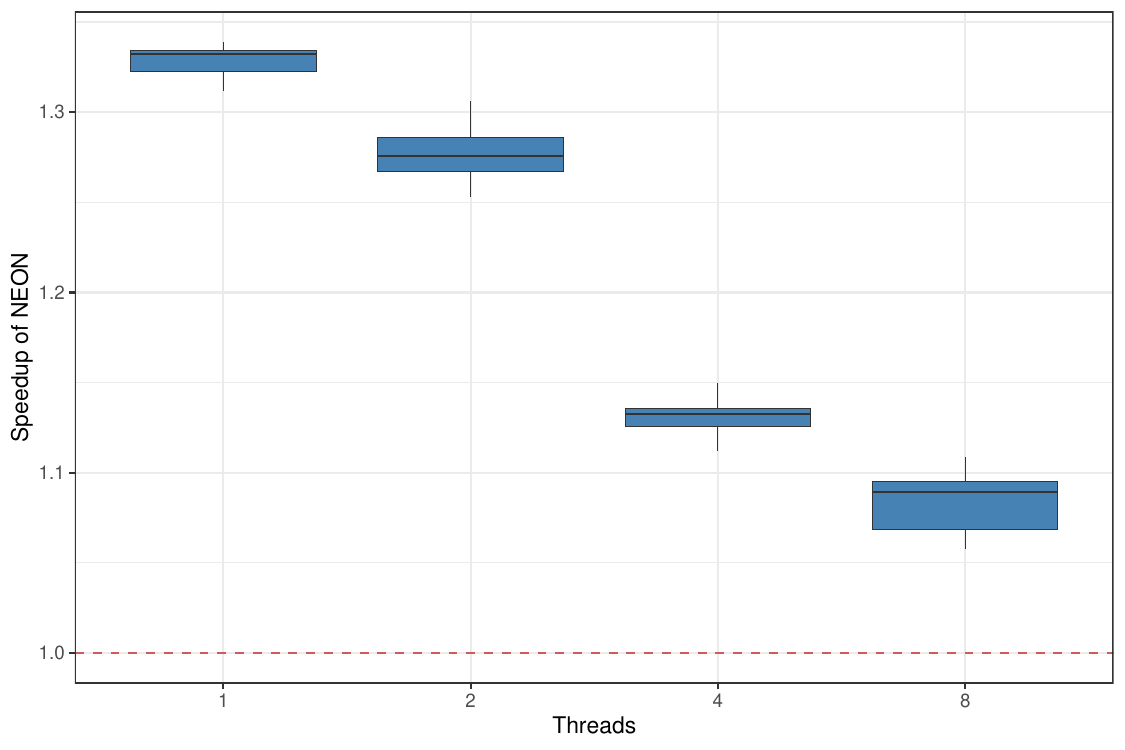}
	\caption{NEON SIMD instructions deliver up to 1.3-fold speedup over non-SIMD implementations on the Apple Inc.\ M3 Pro ARM processor for a nucleotide substitution model with a state-space size of 4.}
	\label{fig:speedup}
\end{figure}

\end{document}